\newcommand{\CLASSINPUTtoptextmargin}{0.75in}%
\newcommand{\CLASSINPUTbottomtextmargin}{1in}%
\newcommand{\CLASSINPUTinnersidemargin}{0.63in}%
\newcommand{\CLASSINPUToutersidemargin}{0.63in}%

\documentclass[conference,10pt,letterpaper]{IEEEtran}%
\usepackage{amsmath}
\usepackage{graphicx}
\usepackage{multirow}
\usepackage[none]{hyphenat}
\usepackage{float}
\usepackage{subfig}
\usepackage{dblfloatfix}
\usepackage{t1enc}
\usepackage{times}
\begin{document}
\raggedbottom
%
%
%
\title{Experimental Study of Periodically Poled Piezoelectric Film Lithium Niobate Resonator at Cryogenic Temperatures}
%
%
%
%
%

\author{\IEEEauthorblockN{Jack Kramer\textsuperscript{1}, Omar Barrera, Sinwoo Cho,
Vakhtang Chulukhadze,  Tzu-Hsuan Hsu, and Ruochen Lu \\}
\IEEEauthorblockA{\textsuperscript{}The University of Texas at Austin, USA
\\ \textsuperscript{1}kramerj99@utexas.edu}
}
\maketitle
%
\begin{abstract}
This work reports the first study of periodically poled piezoelectric film  (P3F) lithium niobate (LiNbO$_3$) resonators at cryogenic temperatures. We experimentally investigate the temperature dependency of resonant frequencies and quality factor ($Q$) of higher-order Lamb modes up to 20 GHz between 80\textdegree K and 297\textdegree  K, using a tri-layer P3F LiNbO$_3$ resonators as the experimental platform. The supported thickness-shear Lamb modes between second-order symmetric (S2) and eleventh-order antisymmetric (A11) modes show temperature coefficients of frequency (TCF) averaging -68.8 ppm/K. Higher $Q$ and more pronounced spurious modes are observed at lower temperatures for many modes. Upon further study, the cryogenic study will be crucial for identifying dominant loss mechanisms and origins of spurious modes in higher-order Lamb wave devices for millimeter-wave applications.  
\end{abstract}
%
%

\section{Introduction}
Microwave acoustic devices have established themselves as the primary method of accomplishing front-end filtering in mobile communication devices \cite{Ruby_Snap}. The phase velocity of sound in piezoelectric solids, such as aluminum nitride (AlN) \cite{GERLICH_AlN_prop, ruby_fbar}, scandium aluminum nitride (ScAlN) \cite{ScAlN_theory,ScAlN_rev}, lithium tantalate (LiTaO$_3$) \cite{litao3_props, sunil_litao} and lithium niobate (LiNbO$_3$) \cite{LN_Temp_Prop, songbin_linbo}, are such that the associated acoustic wavelengths at telecommunication frequencies are on the order of micrometers. Since resonant structures are determined by this characteristic wavelength, this allows for compact acoustic filters to be fabricated that are many orders of magnitude smaller than the corresponding electromagnetic domain devices. However, when these devices are scaled to higher frequencies, this short wavelength becomes an inhibitor for device performance, since the critical dimension of the acoustic resonators scales proportionally, resulting in ultra-thin piezoelectric films \cite{zach_55GHz, Sinwoo_fbar} or small fabrication features \cite{Matteo_2009, Giribaldi2024}. Both solutions pose a significant challenge, as small films suffer from low energy confinement resulting in degraded quality factors ($Q$) and ultra-fine features become unfeasible to fabricate. Additionally, for wide-band filters to be demonstrated, the acoustic electromechanical coupling ($k^2$) should also be maintained. To address these challenges, an interest has developed in so-called periodically poled piezoelectric films (P3F), which refers to a piezoelectric film stack that employs alternating crystal orientations of material. This approach enables a thicker overall structure to maintain $Q$, while simultaneously decreasing the effective critical dimension to scale the device to higher frequencies. The alternating layer stack consequently supports an acoustic mode that is determined by the number of layers employed, and further frequency scaling can be accomplished by utilizing the higher order modes beyond this. Demonstration of this topology has been performed in ScAlN \cite{AlScN_poling_Troy, Acoustis_IUS, acoustis_IMS, k-Band-P3F_troy} and LiNbO$_3$ \cite{Kramer_IFCS, Lu_comp_or}, allowing for a significant increase in the performance of these devices at millimeter-wave (mm-Wave). However, these platforms are still in development, with significant work remaining to properly characterize the current performance limitations. \par

\begin{figure}[t]
\centerline{\includegraphics[width=0.5\textwidth]{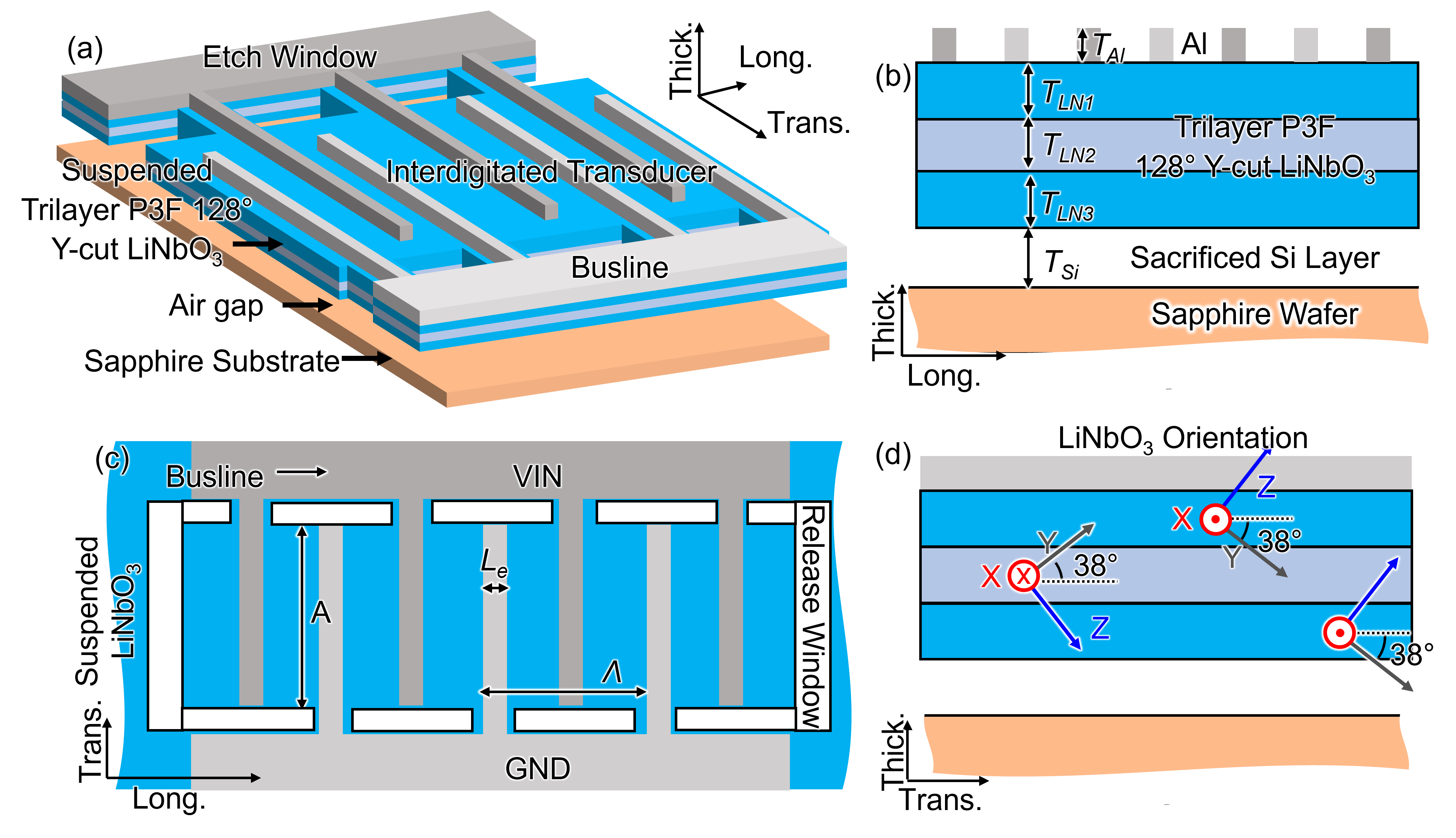}}
\caption{(a) An illustration of the proposed design structure. (b) Transverse cross section of the resonator topology. (b) Top view of the interdigitated transducer structure and parameters. (d) Longitudinal cross section of the film stack, indicating the LiNbO$_3$ orientations of each layer.}
\label{fig:dev_sch}
\end{figure}

For further development of the P3F platform towards filter applications, additional considerations such as temperature dependence must be explored. One important parameter for acoustic design is the temperature coefficient of frequency (TCF). This has been explored extensively in lower frequency surface acoustic wave devices \cite{Guo_2023, Loncar_saw, LN_onSiO2_baw}, but exploration beyond 10 GHz has yet to be performed. This work presents a study of a tri-layer P3F LiNbO$_3$ film stack to explore the dependence of performance versus temperature and to extract the associated TCF for high order modes up to 20 GHz.  

\section{Design and Simulation}
A schematic illustration of the P3F resonator is shown in Fig. \ref{fig:dev_sch} (a)-(d). A tri-layer 128\textdegree Y-cut LiNbO$_3$ film stack is selected, with alternating layers having their material X axis oriented with the -X axis of the adjacent layer, as illustrated in Fig. \ref{fig:dev_sch} (d). The individual LiNbO$_3$ layer thicknesses are designed to be 370 nm, for a total stack thickness of approximately 1.1 $\mu$m. An intermediate amorphous silicon layer (a-Si) is deposited on the LiNbO$_3$ prior to bonding to the sapphire substrate, making the bonding surface the a-Si and sapphire interface. An interdigitated transducer (IDT) on top of the LiNbO$_3$ stack allows for the generation of acoustic lamb waves. Important geometric design considerations for the resonator structure are highlighted in Fig. \ref{fig:dev_sch} (b) and (c). For the set of measured devices, the IDT pitch ($\Lambda$) was set 20 $\mu$m and 24 $\mu$m. The aperture (A) was set to 20 $\mu$m and 30 $\mu$m, while the number of IDT pairs ranged from 4 to 6. The thickness of aluminum (T$_{Al}$) is chosen to be 350 nm to help reduce the series resistance associated with the IDT electrodes. Finally, the width of IDT fingers among the groups ranged from 3 to 6 $\mu$m. \par

\begin{figure}[t]
\centerline{\includegraphics[width=0.5\textwidth]{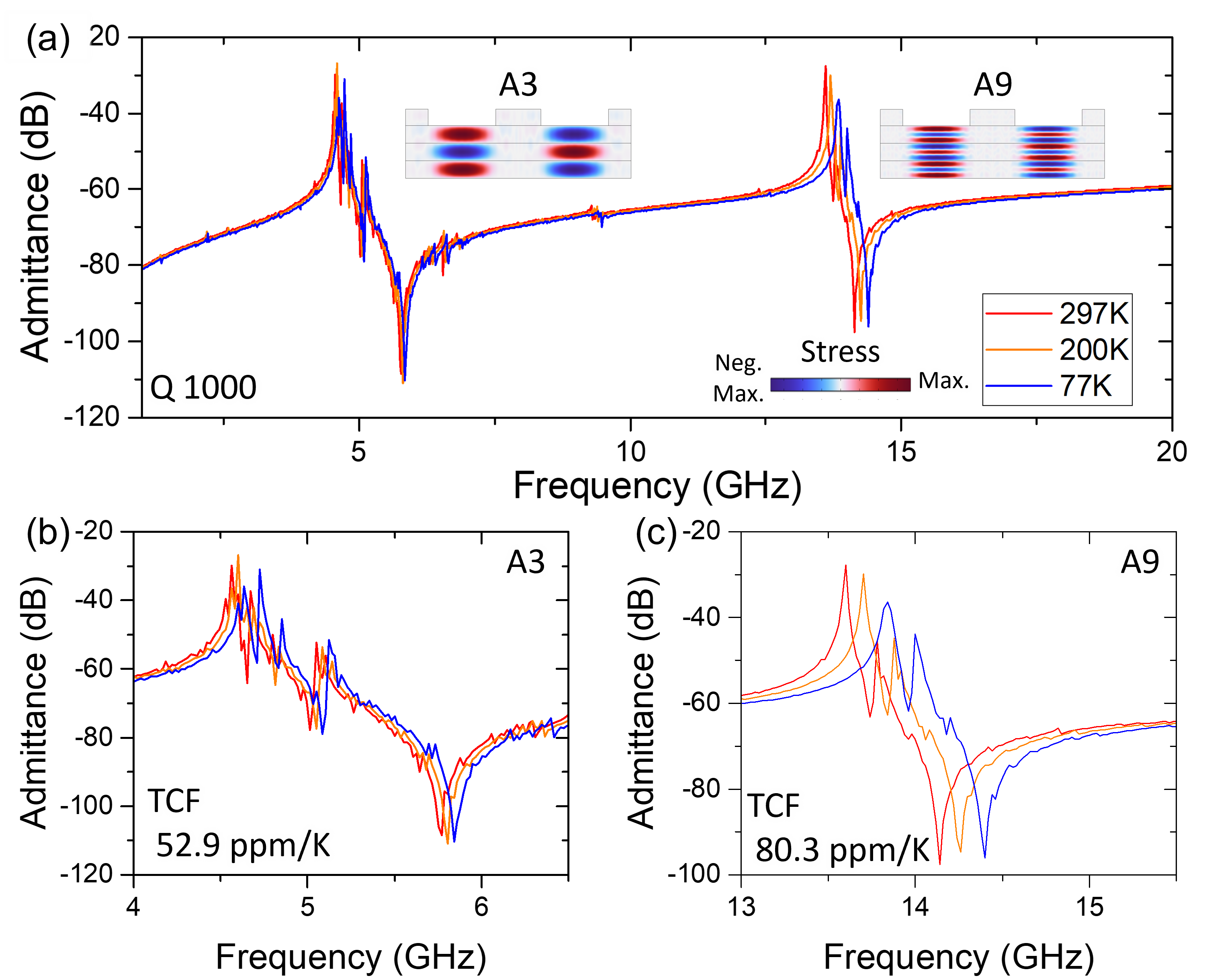}}
\caption{(a) Simulated frequency response of the proposed P3F stack and electrodes for a temperature range from room temperature to liquid nitrogen. The stress mode profiles of the A3 and A9 modes shown adjacent to the associated mode, with the mode profile extracted at parallel resonance. (b)-(c) Zoomed in frequency responses of the respective A3 and A9 modes for the indicated temperatures.}
\label{fig:sim}
\end{figure}

The proposed design was first simulated in COMSOL multiphysics to estimate the TCF and effect on mode profile versus temperature. The simulated frequency response is calculated to 20 GHz, shown in Fig. \ref{fig:sim} (a). For the simulation, the designed LiNbO$_3$ layers with equal thicknesses of 370 nm are assumed for the tri-layer stack. The supported modes within the stack are high order overtones of the first-order antisymmetric lamb mode or A1 mode. The nomenclature for referring to these higher order modes uses either A$N$ or S$N$, where $N$ is the mode order, A indicates an antisymmetric mode and S indicates a symmetric mode in the associated stress profile. In this case, the tri-layer P3F configuration primarily supports the A3 and A9 modes, while other modes are largely suppressed due to cancellation in electric field associated with the displacement of the individual layers. These modes are shown in Fig. \ref{fig:sim} (a), with enlarged plots of the admittance responses plotted in Fig. \ref{fig:sim} (b) and (c). The extracted $k^2$ value is extracted using a fitting of the simulated mode profile and the equation:
\begin{equation}
    k^2 = \frac{\pi ^2}{8} \left( \left( \frac{f_p}{f_s} \right)^2 - 1\right)
\end{equation}  \par
This yields values of 60\% for the A3 mode and 7\% for the A9 mode. To account for temperature dependency and extract the TCF of the mode, elastic stiffness temperature coefficients from \cite{LN_Temp_Prop} and the matrix is rotated to match the 128\textdegree Y-cut LiNbO$_3$. From extracting a linear fitting of the parallel resonant frequency ($f_p$) at each temperature, the TCF can be extracted using:
\begin{equation}
    \text{TCF} = \frac{1}{f_{0,p}} \frac{df_p}{dT} \quad [K^{-1}]
\label{eq_TCF}
\end{equation} 
where $f_{0,p}$ is the y-intercept of the linear fitting and $\frac{df_p}{dT}$ is the slope \cite{MH_Li_TCF}. While the TCF is not exclusive to the parallel resonant peak, in this case the parallel peak is used to make the calculation invariant to the spurious modes that are present in the resonator. Applying this process to the simulated structure yields TCF values of 52.9 ppm/K for the A3 mode and 80.3 ppm/K for the A9 mode.

\section{Fabrication and Measurement}
\begin{figure}[t]
\centerline{\includegraphics[width=0.5\textwidth]{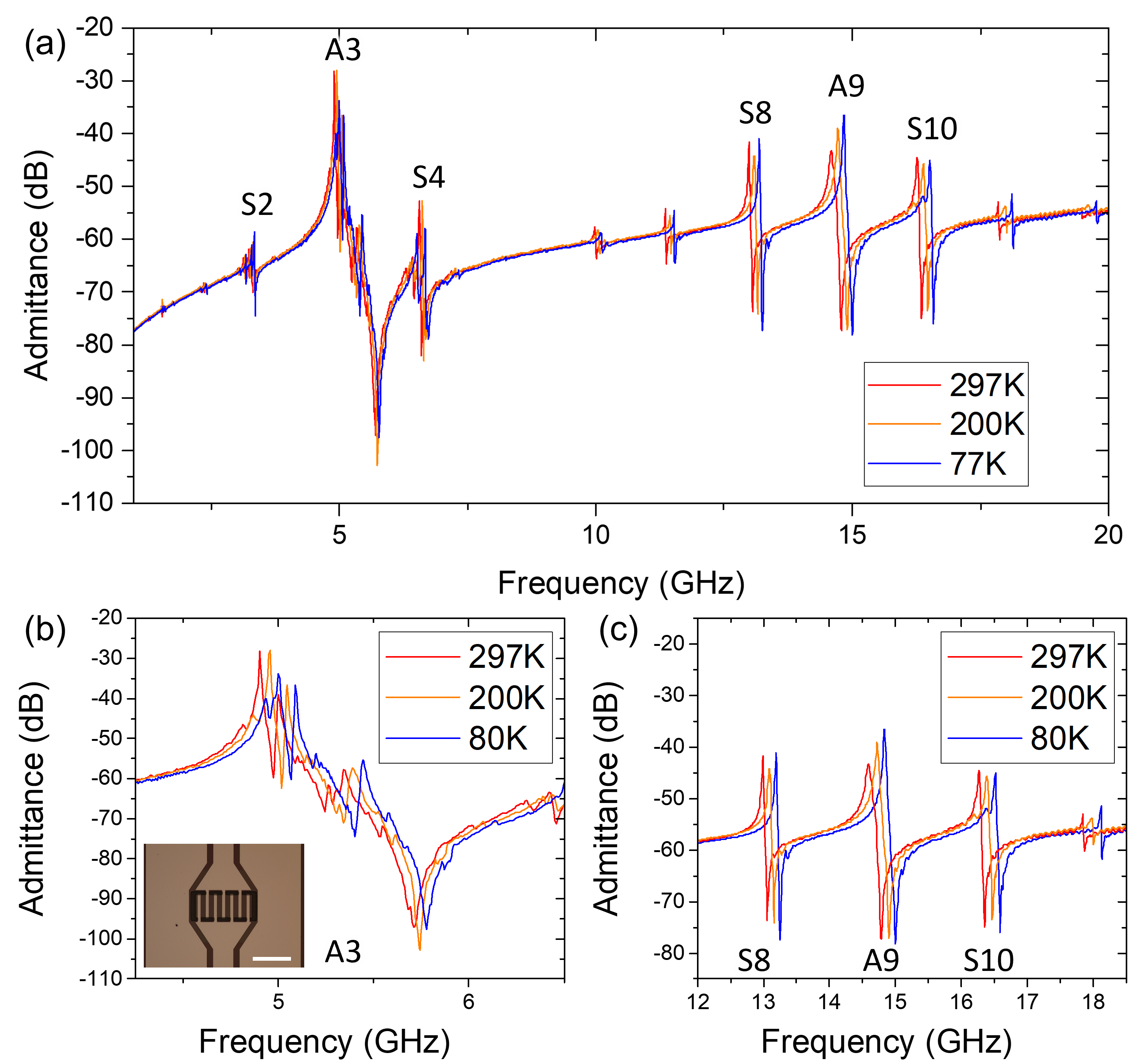}}
\caption{(a) Measured frequency response from 1 to 20 GHz, with the various excited acoustic modes labeled. (b) Zoomed in frequency resonances for A3 and optical image of the measured device. The white bar in the optical image represents a length of 50 $\mu$m. (c) Zoomed in frequency response of the A9 mode, as well as the adjacent S8 and S10 modes.}
\label{fig:meas}
\end{figure}

The fabricated resonator structure is shown in Fig. \ref{fig:meas} (b). The device region is first defined by patterning the device etch windows with photo-lithography and deep oxide etching. At this stage, an over etch into the intermediate silicon layer is required to prepare the device for the later release step. Next, the 350 nm electrode and busline features are patterned using e-beam lithography and evaporation. The unpatterned regions are then removed using a lift-off process. Finally, the device region is suspended from the substrate using gaseous XeF$_2$ etching. \par
The device was measured with use of a Lakeshore TTPX cryogenic probe station and a Keysight P5028A vector network analyzer (VNA). The sample is first mounted to the sample holder with copper tape and vacuum is pulled on the chamber. An initial measurement is then performed at 297K, slightly above the ambient temperature of the room to ensure temperature stability over the course of the measurement. Then, liquid nitrogen (LN$_2$) is used to cool the chamber to 200K and then 80K until the system temperature is stabilized. It should be noted that the stage temperature sensor is used, which may vary slightly compared to the sample surface depending on the thermal conductivity. The number of temperatures at which measurements were performed was limited due to the damage inflicted on the probing pads during the course of each measurement. This damage resulted in poor contact and inconsistent measurement after being probed 4 to 5 times, making a large number of temperatures impractical. At each temperature, the system was calibrated using a commercial short-open-load-thru calibration kit which was mounted to the sample holder adjacent to the test sample. Finally, the device s-parameters are measured on a group of 12 acoustic resonators from 1 to 20 GHz.  

\section{Results and Discussion}
The measured frequency response of a representative device at the three frequencies is shown in Fig. \ref{fig:meas} (a), with zoomed in plots of the A3 and A9 mode regions shown in (b) and (c). When compared to the simulated performance, many additional modes are clearly present. This is due to a mismatch in the individual layer thicknesses that results in intermediate modes that would not be supported in the ideal film structure with equal layer thickness. The extracted parameters for various mode orders are shown in Table \ref{tab1} for a subset of stronger resonances in the device. For many of the additional modes supported in the stack, the associated resonant peak is too small in magnitude to extract accurate performance parameters and are therefore excluded from the table. The clear temperature dependence can be observed, with the expected linear shift in $f_p$ being observed. Additional changes in quality factor can also be observed, although the trend with temperature varies with mode order. This may be due to various factors, including spurious mode temperature dependence affecting the series resonance or the mismatch in thermal coefficients of expansion between aluminum and LiNbO$_3$ resulting in strain that enhances $Q_s$ at specific temperatures. Additionally, many spurious modes present in the film stack show enhanced $Q$ values and corresponding frequency shifts. Since many of these spurious modes are present in the simulation as well and correspond to electrode motion, this could become a tool for correlating various loss contributors. Thus, the general increase in quality factor for many of the main modes at cryogenic temperatures can be used as important data points in the future extraction of fundamental loss mechanisms for P3F films.  \par

\begin{figure}[t]
\centerline{\includegraphics[width=0.5\textwidth]{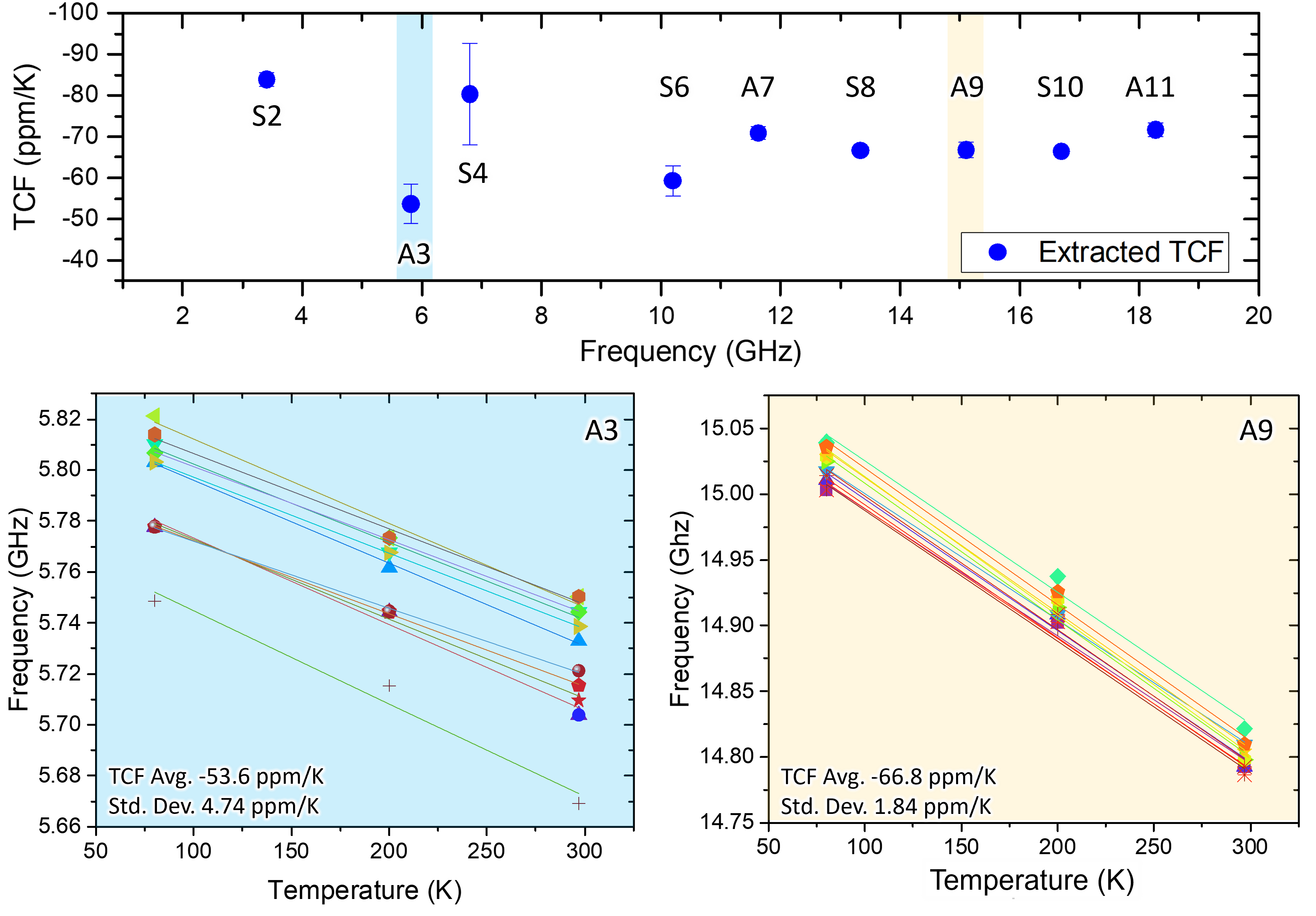}}
\caption{(a) Extracted average parallel resonance TCF values for prominent modes from 1 to 20 GHz from 12 different resonators. Error bars represent one standard deviation. (b)-(c) Linear fittings of parallel resonant frequency as a function of temperature for A3 and A9 modes, respectively.}
\label{fig:TCF_update}
\end{figure}

The extracted average TCF values for the selected group of devices are plotted in Fig. \ref{fig:TCF_update} (a) for the well defined modes. Although the signal for some of the minor modes is insufficient for performance parameter extraction, the associated $f_p$ is well defined. Thus, at each mode the associated $f_p$ is extracted versus temperature. Then a linear fitting for each device is performed, as shown for the A3 and A9 modes in Fig. \ref{fig:TCF_update} (b) and (c). From this, the associated slopes and intercepts can be used to calculate the TCF of the device using equation (\ref{eq_TCF}). The average value and standard deviation of the TCF values are then extracted for each mode. While the TCF associated with each mode does vary, for the high order modes surrounding A9, the TCF values converge towards the total average value of -68.5 ppm/K. Since these high order modes are closer to the pure lamb mode, with fewer spurious modes, it is reasonable to conclude that the TCF of the stack is consistent across mode order for this frequency range. 
\begin{table}[t]
\caption{Resonance Summary}
\begin{center}
\begin{tabular}{|c|c|c|c|c|}
\hline
\cline{2-4} 
\textbf{Mode} & \textbf{Parameter (unit)}  & \textbf{\textit{297K}} & \textbf{\textit{200K}} & \textbf{\textit{80K}} \\
\hline
S2 & $f_p$ (GHz) & 3.29 & 3.32 & 3.35 \\
\cline{2-5}
 & $Q_s$ & 301 & 461 & 421 \\
\hline

A3 & $f_p$ (GHz) & 4.90 & 4.95 & 4.99 \\
\cline{2-5}
 & $Q_s$ & 260 & 340 & 170 \\
 \hline
 
S4 & $f_p$ (GHz) & 6.56 & 6.62 & 6.68 \\
\cline{2-5}
 & $Q_s$ & 380 & 640 & 520 \\
\hline

\hline
S8 & $f_p$ (GHz) & 12.99 & 13.10 & 13.19 \\
\cline{2-5}
 & $Q_s$ & 640 & 640 & 760 \\

\hline
A9 & $f_p$ (GHz) & 14.62 & 14.73 & 14.84 \\
\cline{2-5}
 & $Q_s$ & 243 & 323 & 500 \\

\hline
S10 & $f_p$ (GHz) & 16.27 & 16.40 & 16.52 \\
\cline{2-5}
 & $Q_s$ & 485 & 385 & 445 \\

\hline
\end{tabular}
\label{tab1}
\end{center}
\end{table}

\section{Conclusion}
This work reports the first study of P3F LiNbO$_3$ resonators up to 20 GHz at cryogenic temperatures between 80 K and 297 K. Using a 1.1$\mu$m thick tri-layer 128\textdegree LiNbO$_3$ resonator as the test platform, we experimentally investigate the TCF and $Q$ of higher-order Lamb modes in P3F stacks up to 20 GHz. TCF of S2 to A11 modes are measured to be around -70 ppm/K, with an average of -68.8 ppm/K. Higher $Q$ and more pronounced spurious modes are observed at lower temperatures, pointing towards a potential method for determining material and electrical loss mechanisms in the resonator. Upon further study, the cryogenic study will be a key method for assisting optimization of P3F resonators for millimeter-wave applications.


\section*{Acknowledgment}
The authors would like to thank the funding of the DARPA COFFEE program for supporting this work.



\bibliographystyle{IEEEtran}

\bibliography{ref}

\end{document}